\documentstyle[preprint,tighten,aps,epsf,floats]{revtex} 

\def\lqcd{\Lambda_{\rm QCD}}
\def\WB{W^{(B,{\rm sl})}}
\def\Wt{W^{(\tau)}}
\def\ot{\Omega^{(\tau)}}
\def\oslB{\Omega^{(B,{\rm sl})}}
\def\gslB{\Gamma^{(B,{\rm sl})}}
\def\gnlB{\Gamma^{(B,{\rm nl})}}

\def\gslb{\Gamma^{(B,{\rm sl})}}
\def\gnlb{\Gamma^{(B,{\rm nl})}}
\def\ght{\Gamma^{(\tau,{\rm h})}}
\def\me#1#2#3{\langle #1 | #2 | #3 \rangle}
\def\ord#1{{\mathcal O}\left( #1 \right)}
\def\abs#1{\left| #1 \right|}
\def\ct{a^{(\tau)}}
\def\csl{a^{(B,{\rm sl})}}
\def\cnl{a^{(B,{\rm nl})}}

\begin{document}

\preprint{\tighten \vbox{\hbox{UCSD/PTH 98-07} \hbox{hep-ph/9803236} \hbox{} }}

\title{Local duality and nonleptonic $B$ decay}
\author{Zoltan Ligeti\footnote{zligeti@ucsd.edu} and 
  Aneesh V.~Manohar\footnote{amanohar@ucsd.edu} }
\address{Department of Physics, University of California at San Diego,\\
9500 Gilman Drive, La Jolla, CA 92093--0319}
\date{March 1998}
\maketitle

\begin{abstract}%
In the large $N_c$ limit, the nonleptonic width of the $B$ meson is determined
by the lepton mass spectrum in semileptonic $B$ decay and the hadron mass
spectrum in $\tau$ decay.  This result can be compared with the nonleptonic
width computed using the operator product expansion and local duality, to
estimate the violation of local duality in $B$ decay. In the absence of the
required data, we use experimental Monte Carlo predictions and find that the
difference is small, at the few percent level. We estimate the theoretical
uncertainty in our results from higher order corrections (including $1/N_c$
effects).  We also identify a new contribution to the $\Lambda_b-B$ lifetime
difference.

\end{abstract}

\newpage
\section{Introduction}

Inclusive decays for heavy-quark hadrons such as $B$ mesons are computed in a
systematic expansion in inverse powers of the heavy quark mass using the
operator product expansion (OPE). Decay rates are related to the discontinuity
of scattering amplitudes at physical cuts, so the calculation is necessarily
performed for time-like momentum transfer and is not a short-distance process
for which the OPE is valid. Nevertheless, one can still use the OPE for
suitably averaged quantities.  The simplest example is $R(s)$, the inclusive
rate for $e^+ e^- \to {\rm hadrons}$, which is related to the discontinuity in
the vacuum polarization amplitude $\Pi(s)$ along the real axis for $s>0$.  The
operator product expansion can be applied to $\Pi(s)$ far away from the
physical cuts, such as at complex $s=s_0 + i \Delta$, $\Delta\gg\Lambda_{\rm
QCD}$, to give $\Pi(s)$ as an expansion in $\lqcd^2/s$ and $\alpha_s$.
Analyticity can then be used to relate $\Pi(s_0+i\Delta)$ to an integral of the
form \cite{PQW}
\begin{equation}
\int_0^\infty {R(s) \over (s-s_0)^2 + \Delta^2}\, {\rm d}s\,. 
\end{equation}
Thus smeared averages of $R(s)$ can be computed using perturbation theory. The
$\alpha_s$ corrections to these depend on the smearing width $\Delta$, and are
not under control as $\Delta \to 0$. In practice $\Delta$ is made as small as
possible while keeping the radiative corrections under control.  Typically
$\Delta$ is chosen to be $\Delta \gtrsim 500\,$MeV. One expects that
perturbative computations agree with data, provided one averages over a large
number of hadronic resonances.\footnote{There are examples where local duality
is proven to hold even though only two exclusive decay modes
contribute~\cite{boyd}.}

A similar situation exists in inclusive semileptonic $B$ decays, $B \to
X_c\,e\,\bar\nu$ \cite{CGG,VoSi}. The differential decay distribution ${\rm
d}\gslB/{\rm d}q^2 {\rm d}E_e {\rm d}E_\nu$ can be computed using the operator
product expansion \cite{BSUV,MaWi,Blok,Mannel}. Here $q^2$ is the invariant
mass of the lepton pair, and $E_e$ and $E_\nu$ are the energy of the electron
and neutrino, respectively.  The total semileptonic decay rate $\gslB$
automatically involves averaging over different hadronic states, and one
expects that the OPE computation for $\gslB$ is valid. Inclusive nonleptonic $B
\to X_c$ decays can also be computed using an OPE \cite{VoSi,inclnl}. In this
case, there is no variable such as $q^2$ to integrate over, so the computation
has no intrinsic smearing. One must then make the assumption of local duality,
that the quark and hadron computations agree at a single kinematic point. In
the case of $R(s)$, this would mean that one could compute $R(s)$ for each
value of $s$. This is clearly false when there are narrow resonances (e.g., in
the $\Upsilon$ region), but is valid elsewhere because the large number of
overlapping hadronic resonances provide a natural smearing mechanism. An
important question is the extent to which local duality holds for nonleptonic
$B$ meson decays. There is a long-standing claimed discrepancy between theory
and experiment in the $B$ meson semileptonic branching ratio and in the
$\Lambda_b$ to $B$ lifetime ratio, and the failure of local duality is one way
to resolve these.

In this paper, we study the violation of local duality using the $1/N_c$
expansion of QCD \cite{tHooft}. In this limit, we show that the nonleptonic $B$
meson decay rate $\gnlB$ can be expressed in terms of the product of ${\rm
d}\gslB/{\rm d}q^2$, the invariant mass distribution of the lepton pair in
semileptonic $B$ decay, and ${\rm d}\ght/{\rm d}q^2$, the hadron mass
distribution in nonleptonic $\tau$ decay.  One gets a quantitative measure of
the violation of local duality in the large $N_c$ limit by comparing the
prediction for $\gnlB$ using the OPE with that using the experimental data for
${\rm d}\gslB/{\rm d}q^2$ and ${\rm d}\ght/{\rm d}q^2$, and a large $N_c$
factorization formula derived here.  We estimate the accuracy to which the
large $N_c$ results should be valid in QCD.

\section{Basic definitions and results}

Nonleptonic $B$ meson decays are dominantly due to the weak
Hamiltonian\footnote{The contribution of $b\to u$ decay to the total
nonleptonic $B$ decay width is small, and we neglect it in this paper. There
are also terms in the weak Hamiltonian for $b\to c \bar c (d+s)$ decay, which
are discussed later.}
\begin{equation}\label{HW}
H =  {4G_F\over\sqrt2}\, \cnl\, V_{cb} V_{uq}^*\, \Big[
  c_1(m_b)\, (\bar c\,\gamma_\mu P_L\, b)\, (\bar q\,\gamma^\mu P_L\, u)  
  + c_2(m_b)\, (\bar q\,\gamma_\mu P_L\, b)\, (\bar c\,\gamma^\mu P_L\, u) 
  \Big]\,,
\end{equation}
where $P_L=(1-\gamma_5)/2$ and $q=d,s$.  The Hamiltonian in Eq.~(\ref{HW}) is
renormalized at the scale $\mu=m_b$. The coefficient  $\cnl = 1 +
(2{\alpha_{\rm em} / 3\pi}) \ln ({M_W / m_b})$  is the electromagnetic
correction. The coefficients $c_i$ are QCD corrections given by $c_{1,2}=(c_+
\pm c_-)/2$, where
\begin{equation}\label{cpcm}
c_+(\mu) = \left[{\alpha_s(\mu) \over \alpha_s(M_W)} \right]^{-6/23},
  \qquad
c_-(\mu) = \left[{\alpha_s(\mu) \over \alpha_s(M_W)} \right]^{12/23},
\end{equation}
at one loop order. At next-to-leading order \cite{BuBu}, $c_1(m_b)=1.13$ and
$c_2(m_b)=-0.29$ for $\alpha_s(M_Z)=0.118$. The $c_1$ and $c_2$ terms are
effective charged and neutral current interactions, respectively.

Nonleptonic $B$ decay widths are computed by taking the decay amplitude
$\me{X_c}{H}{B}$, squaring, and summing over all possible final states. A
simpler method is to compute the imaginary part of the $B \to B$ forward
scattering amplitude with an insertion of $H$ and $H^\dagger$. Diagrams
involving spectator quarks \cite{GNPR,BGR}, such as Pauli interference, weak
annihilation, and $W$-exchange are of order $\lqcd^3/m_b^3$ relative to free
quark decay (since they depend on matrix elements of four-quark operators), and
can be neglected for the purposes of this paper. In the large $N_c$ limit,
annihilation effects are enhanced by a factor $N_c$ relative to free quark
decay. While this means that annihilation diagrams are formally dominant at
$N_c = \infty$, they can still be neglected for $N_c=3$ because the $N_c$
enhancement does not compensate for the $1/m_b^5$ suppression factor, since the
diagram is helicity suppressed and involves four-quark operators.

\begin{figure}
\centerline{\epsfysize=5cm\epsffile{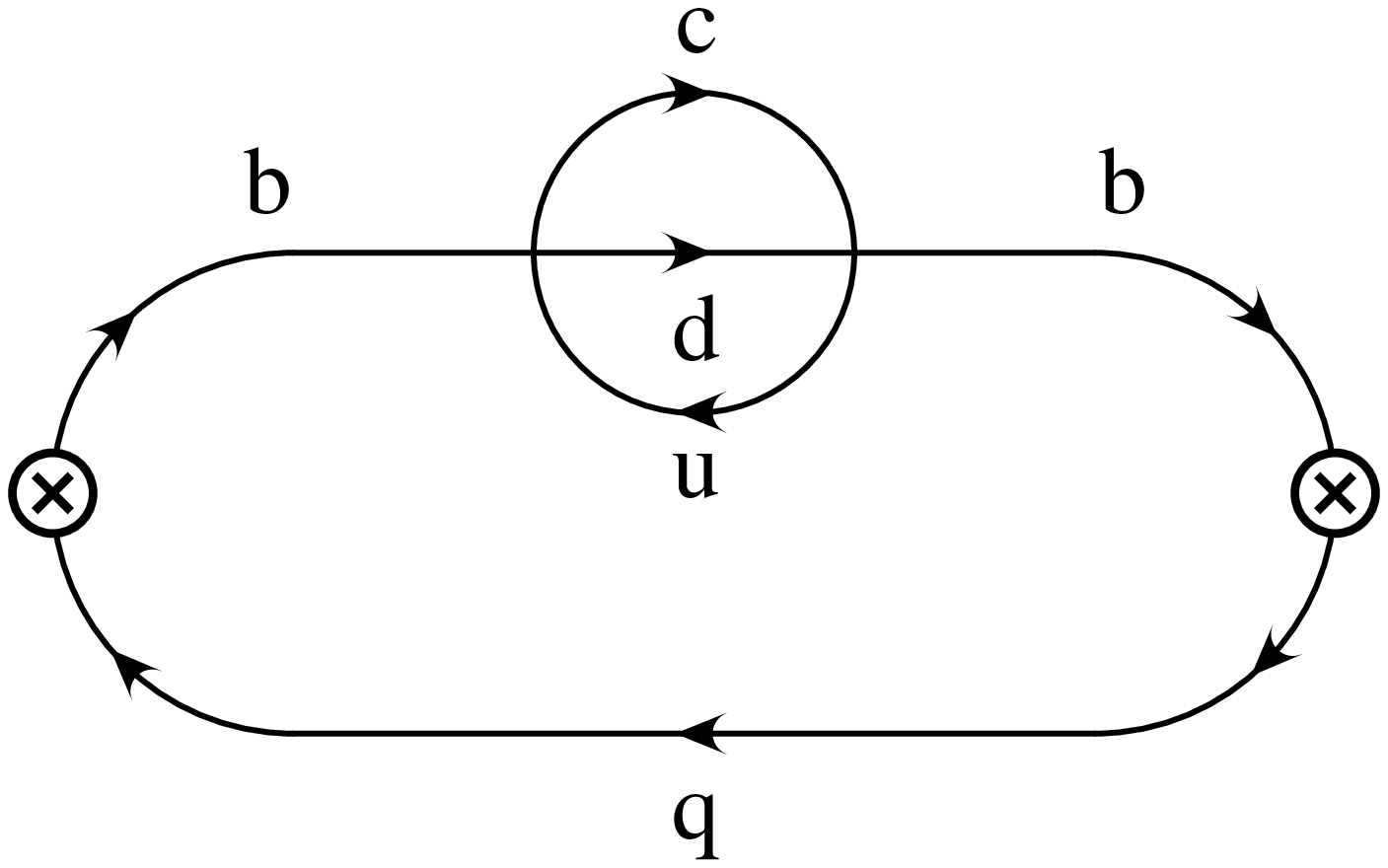}}
\caption{Leading contribution to $B$ meson decay ($B\to B$ forward scattering
amplitude).  The weak currents have been contracted to a point, and $\otimes$
represent quark bilinears that create or annihilate $B$ mesons. 
\label{fig:1}}
\end{figure}

\begin{figure}
\centerline{\epsfysize=10cm\epsffile{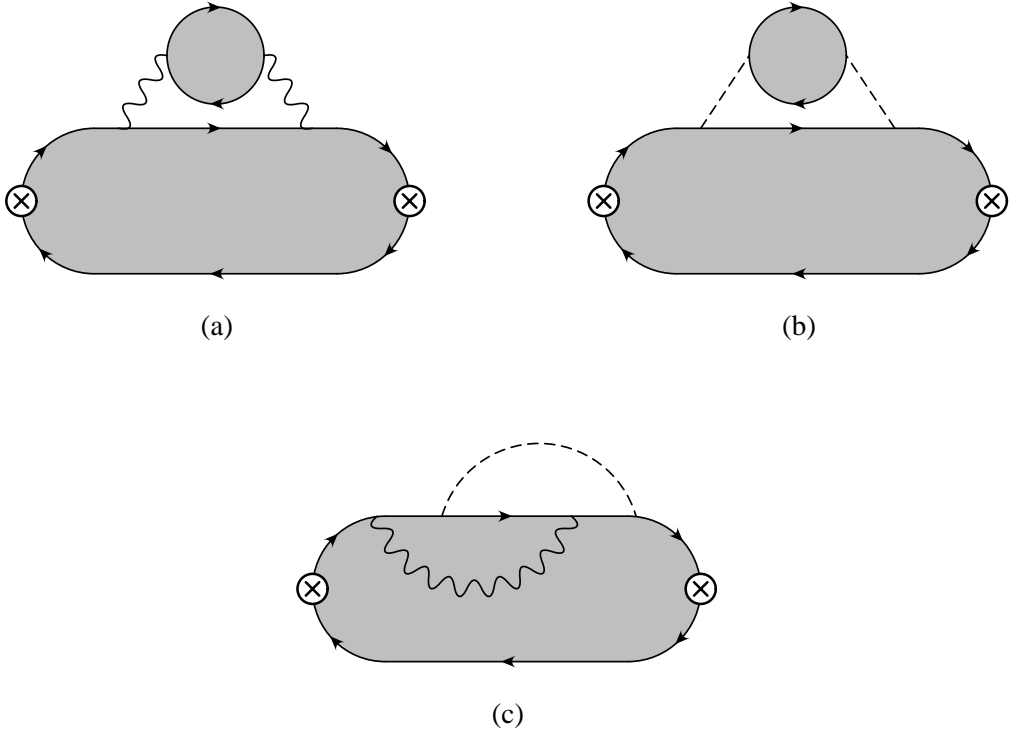}}
\caption{Free quark decay diagrams showing the color flow. The charged
and neutral current interactions in Eq.~(\ref{HW}) are represented by the
exchange of gauge bosons denoted by wiggly and dashed lines, respectively.
The shaded regions represent arbitrary planar gluons.
\label{fig:2}}
\end{figure}

The free quark decay diagram in Fig.~\ref{fig:1} is redrawn in Fig.~\ref{fig:2}
to show the color flows.  The charged and neutral current terms in the weak
Hamiltonian in Eq.~(\ref{HW}) are denoted by the exchange of color singlet
charged and neutral ``gauge bosons.'' This is a convenient way of showing the
color flows in the diagrams. The order in $N_c$ of the various graphs can be
computed by using a factor of $N_c$ for each closed color loop, and
$1/\sqrt{N_c}$ for each external meson. In the large $N_c$ limit,
Fig.~\ref{fig:2}(a) and (b) are $\ord{N_c}$, and (c) is $\ord 1$.
Fig.~\ref{fig:2}(a), (b) and (c) will be referred to as the charged current,
neutral current, and interference diagrams, respectively.  In the large $N_c$
limit, arbitrary planar gluons must be summed over, which is represented
schematically by the shaded regions in the figures.  The leading diagrams have
arbitrary planar exchanges within each quark loop, but exchanges between the
two quark loops are suppressed by at least two powers of $N_c$. Thus in the
large $N_c$ limit, Fig.~\ref{fig:2}(a) and (b) factorize into the product of
two non-perturbative matrix elements.  They are: (i) The matrix element of two
currents in a $B$ meson; (ii) The matrix element of two currents in the vacuum.
These will be related to inclusive semileptonic $B$ decay and hadronic $\tau$
decay, respectively.

Consider the inclusive decay of an initial particle $I$ with velocity $v$ due
to the Hamiltonian 
\begin{equation}\label{HW1}
H = {4G_F\over\sqrt2}\, J^\mu j_\mu \,.
\end{equation}
The decaying particle $I$ and the weak currents $J^\mu$ and $j^\mu$ will be
different for the three cases we need in this paper---nonleptonic $B$ decay,
semileptonic $B$ decay, and hadronic $\tau$ decay. In these decays, the weak
Hamiltonian has factors in addition to those shown explicitly in
Eq.~(\ref{HW1}), such as CKM angles or renormalization group coefficients,
which will be included later.

Define the tensors
\begin{equation}\label{Wmunu}
W^{\mu\nu}(v,q) = (2\pi)^3\, \sum_X \delta^4(m_I\, v - q - p_X)\,
  \langle I(v)| J^{\mu\dagger} | X \rangle\,
  \langle X| J^\nu | I(v) \rangle \,,
\end{equation}
and
\begin{equation}\label{Omunu}
\Omega^{\mu\nu}(q) = (2\pi)^4\, \sum_Y \delta^4(q-p_Y)\,
  \langle 0| j^{\mu\dagger} | Y \rangle\,
  \langle Y| j^\nu | 0 \rangle \,.
\end{equation}
In terms of $W$ and $\Omega$, the weak decay rate is given by
\begin{equation}\label{rate}
{{\rm d}\Gamma\over {\rm d}q^2\,{\rm d}q_0} = 
  {2G_F^2\over\pi^2} \sqrt{q_0^2-q^2}\, 
  W^{\mu\nu}(v,q)\, \Omega_{\mu\nu}(q) \,,
\end{equation}
as long as the diagrams for the process factorize, e.g., as in
Fig.~\ref{fig:2}(a) or (b). This is the factorization formula which we will use
in the rest of the paper.

The hadronic tensor $W^{\mu\nu}$ is identical to the one that appears in the
OPE for semileptonic decays \cite{MaWi,Blok}.  It can be expanded as
\begin{equation}\label{Wdecompose}
W^{\mu\nu} = -g^{\mu\nu} W_1 + v^\mu v^\nu W_2 
  - i \varepsilon^{\mu\nu\alpha\beta} v_\alpha q_\beta W_3
  + q^\mu q^\nu W_4 + (q^\mu v^\nu + q^\nu v^\mu) W_5 \,,
\end{equation}
where $W_i$ are functions of $q^2$ and $q\cdot v$.
$\Omega^{\mu\nu}$ can be decomposed into transverse and longitudinal parts
\begin{eqnarray}\label{Odecompose}
\Omega^{\mu\nu} = (-q^2 g^{\mu\nu} + q^\mu q^\nu)\, \Omega_T 
  + q^\mu q^\nu\, \Omega_L \,,
\end{eqnarray}
where $\Omega_{T,L}$ are functions of $q^2$.
Contracting Eqs.~(\ref{Wdecompose}) and (\ref{Odecompose}) gives
\begin{eqnarray}
W^{\mu\nu}\, \Omega_{\mu\nu} &=& 
  \Omega_T \left[ 3 q^2\, W_1 - q^2\, W_2 + (q\cdot v)^2\, W_2 \right] 
  \nonumber\\*
&+& \Omega_L \left[ -q^2\, W_1 + (q\cdot v)^2\, W_2 + q^4\, W_4 
  + 2 q^2 (q\cdot v)\, W_5 \right] . \label{womeg}
\end{eqnarray}

Both $W^{\mu\nu}$ and $\Omega_{\mu\nu}$ are calculable in an operator product
expansion.  They are related to the discontinuities across the cut of the 
amplitudes $T^{\mu\nu}$ and $\Pi_{\mu\nu}$ defined by
\begin{eqnarray}\label{TPidef}
T^{\mu\nu} &=& -i \int {\rm d}^4x\, e^{-iq\cdot x}\,
  \langle B(v) |\, T\{ J^{\mu\dagger}(x)\, J^\nu(0)\} | B(v) \rangle \,,
  \nonumber\\*
\Pi^{\mu\nu} &=& -i \int {\rm d}^4x\, e^{-iq\cdot x}\,
  \langle 0 |\, T\{ j^{\mu\dagger}(x)\, j^\nu(0)\} | 0 \rangle \,,
\end{eqnarray}
i.e., $W^{\mu\nu}={\rm Im}\,T^{\mu\nu}/\pi$ and $\Omega^{\mu\nu}=2\,{\rm
Im}\,\Pi^{\mu\nu}$.  $T^{\mu\nu}$ and $\Pi^{\mu\nu}$ have a form factor
decomposition similar to $W^{\mu\nu}$ and $\Omega^{\mu\nu}$ in
Eqs.~(\ref{Wdecompose}) and (\ref{Odecompose}), with $W_i \to T_i$ and
$\Omega_i \to \Pi_i$.  In the complex $q^2$ plane, $\Pi^{\mu\nu}$ has a cut
along the real axis for $q^2\gtrsim0$.  $T^{\mu\nu}$ has two cuts along the
real axis for $q^2\lesssim(m_b-m_c)^2$ corresponding to intermediate states
with a charm quark, and for $q^2\gtrsim(m_b+2m_c)^2$ corresponding to
intermediate states with a $\bar c$ and two $b$ quarks.

\subsection{Inclusive hadronic $\tau$ decay}

The weak Hamiltonian at $\mu=m_\tau$ for hadronic $\tau$ decay 
$\tau \to \nu_\tau\,X$ is
\begin{equation}\label{HWtau}
H = {4G_F\over\sqrt2}\, \ct\, V_{uq}^*\, 
  (\bar \nu_\tau\,\gamma_\mu P_L\, \tau)\, (\bar q\,\gamma^\mu P_L\, u).
\end{equation}
$\ct = 1 + (\alpha_{\rm em} / \pi) \ln (M_W / m_\tau)$  is the electromagnetic
correction. There are no QCD corrections to Eq.~(\ref{HWtau}) since the quark
part is a  conserved current.

The hadronic $\tau$ decay amplitude clearly factors into hadronic and leptonic
parts, irrespective of whether $N_c$ is large or not, since no gluons can be
exchanged between the quark and lepton parts of the diagram. The leptonic part
of the diagram is calculable, and gives
\begin{equation}
\begin{array}{ll}
\Wt_1 = \frac12\,(m_\tau-q_0)\,\delta(\Delta)\,,\qquad &
\Wt_2 = m_\tau\,\delta(\Delta)\,, \\
\Wt_3 = -\Wt_5 = \frac12\,\delta(\Delta)\,, &
\Wt_4 = 0\,,
\end{array}
\end{equation}
where $\Delta=m_\tau^2+q^2-2m_\tau q_0$. The hadronic part of the $\tau$ decay
diagram is precisely $\Omega$ for the charged current nonleptonic $B$ decay
diagram. The factorization formula Eq.~(\ref{rate}) can be applied to this
case, with $W_{\mu\nu}$ set equal to $\Wt_{\mu\nu}$. $\Omega$ in $\tau$ decay
is defined to be $\ot_{T,L}$, the sum of $\Omega$'s for the weak currents
$V_{ud}(\bar d\,\gamma_\mu P_L\,u)$ and $V_{us}(\bar s\,\gamma_\mu P_L\,u)$, so
that the CKM angle is included in $\ot$.  Integrating over ${\rm d}q_0$ using
$\delta(\Delta)$, gives \cite{BNP}
\begin{equation}\label{tauspec}
{{\rm d}\ght \over {\rm d}q^2} = {G_F^2\over 8\pi^2\, m_\tau^3} \abs{\ct}^2 
  (m_\tau^2-q^2)^2 \left[(m_\tau^2+2q^2)\,\ot_T + m_\tau^2\,\ot_L\right] ,
\end{equation}
where $q^2$ is the invariant mass-squared of the hadrons in the final state.

\subsection{Inclusive semileptonic $B$ decay}

Inclusive semileptonic $B\to X_c\,e\,\bar\nu$ decay is due to the weak
Hamiltonian
\begin{equation}\label{HWbsemi}
H = {4G_F\over\sqrt2}\, \csl\, V_{cb} \, (\bar c\,\gamma_\mu P_L\, b)\,
  (\bar e \,\gamma^\mu P_L\, \nu).
\end{equation}
$\csl = 1 + (\alpha_{\rm em}/\pi) \ln(M_W/m_b)$ is the electromagnetic 
correction.  There are no QCD corrections since the quark part of the
Hamiltonian is a conserved current.  As for hadronic $\tau$ decay, the
semileptonic $B$ decay diagram factorizes into a hadronic and leptonic part,
irrespective of whether $N_c$ is large or not. The leptonic part
$\Omega^{\mu\nu}$ is calculable,
\begin{eqnarray}\label{OmegaEM}
\oslB_T &=& {1\over12\pi}\, 
  {(q^2-m_e^2)^2\,(2q^2+m_e^2)\over 2q^6} \,, \nonumber\\*
\oslB_L &=& {1\over8\pi}\, 
  {(q^2-m_e^2)^2\,m_e^2\over q^6} \,,
\end{eqnarray}
where $q^2$ is the invariant mass-squared of the lepton pair.  Therefore, ${\rm
d}\gslB/{\rm d}q^2{\rm d}q_0$ gives information on the hadronic part
$\WB_{\mu\nu}$.  Using differential $B$ decay spectra for $e$, $\mu$, and
$\tau$, it is possible to measure $\WB_{1,\ldots,5}$ \cite{FLNN}.  This seems
to be a remote possibility, so in the rest of this paper we neglect lepton
masses. In the $m_e\to0$ limit, Eqs.~(\ref{rate}) and
(\ref{Wdecompose}) give
\begin{equation}\label{slrate}
{{\rm d}\gslB \over {\rm d}q^2\,{\rm d}q_0} = {2G_F^2\over\pi^2} \abs{\csl}^2
  \sqrt{q_0^2-q^2} \left[3q^2\, \WB_1 + (q_0^2-q^2)\, \WB_2 \right]
  {1 \over 12 \pi} \,,
\end{equation}
where the CKM angle $\abs{V_{cb}}^2$ has been absorbed into the definition of
$\WB_{1,2}$.

\subsection{The OPE calculation for nonleptonic $B$ decay}

The nonleptonic $B$ decay rate can be computed using perturbation theory if
one assumes local duality.  To order $\Lambda_{\rm QCD}^2/m_b^2$,  the result
is \cite{inclnl}
\begin{eqnarray}
\gnlb [b \to c \bar u (d+s)] = 3 \Gamma_0 \abs{\cnl}^2 &&{} 
  \bigg\{ {c_-^2+2c_+^2\over3}\, \bigg[ \bigg( 1 + {\lambda_1\over2m_b^2} \bigg)
  + {3\lambda_2\over m_b^2} \bigg( \rho\,{{\rm d}\over{\rm d}\rho} - 2 \bigg)
  \bigg] f(\rho) \nonumber\\*
&&{} + 4(c_-^2-c_+^2)\, {\lambda_2\over m_b^2}\, (1-\rho)^3
  \bigg\} + \ldots \,, \label{bcu} \\
\gnlb [b \to c \bar c (d+s)] = 3 \Gamma_0 \abs{\cnl}^2 &&{} 
  \bigg\{ {c_-^2+2c_+^2\over3}\, \bigg[ \bigg( 1 + {\lambda_1\over2m_b^2} \bigg)
  + {3\lambda_2\over m_b^2} \bigg( \rho\,{{\rm d}\over{\rm d}\rho} - 2 \bigg)
  \bigg] g(\rho) \nonumber\\*
&&{} + 4(c_-^2-c_+^2)\, {\lambda_2\over m_b^2}\, h(\rho)
  \bigg\} + \ldots \,,\label{bcc}
\end{eqnarray}
where $\rho=m_c^2/m_b^2$, $\Gamma_0=G_F^2 m_b^5 |V_{cb}|^2/(192\pi^3)$, and
\begin{eqnarray}
f(\rho) &=& 1-8\rho+8\rho^3-\rho^4-12\rho^2\ln\rho \,,\nonumber\\*
g(\rho) &=& \sqrt{1-4\rho}\,(1-14\rho-2\rho^2-12\rho^3)
  + 24\rho^2(1-\rho^2) \ln{1+\sqrt{1-4\rho}\over1-\sqrt{1-4\rho}} \,,
  \nonumber\\*
h(\rho) &=& \sqrt{1-4\rho}\,(1+\rho/2+3\rho^2)
  - 3\rho(1-2\rho^2) \ln{1+\sqrt{1-4\rho}\over1-\sqrt{1-4\rho}} \,.
\end{eqnarray}
The ellipses in Eqs.~(\ref{bcu}) and (\ref{bcc}) denote corrections of order
$\Lambda_{\rm QCD}^3/m_b^3$ and $\alpha_s(m_b)$. The total nonleptonic decay
rate is the sum of Eq.~(\ref{bcu}) and Eq.~(\ref{bcc}). $\alpha_s$ corrections
will be included when we use these results in the next section.

\section{A (not so) toy model: $\lowercase{c}_2=0$}

The total nonleptonic decay rate depends on the short distance QCD corrections
$c_1$ and $c_2$. The validity of local duality depends on long distance
effects, such as the masses and widths of hadronic resonances. One can test
local duality in the nonleptonic $B$ decay rate by studying a hypothetical
world in which $c_2=0$ and $c \bar c (d+s)$ final states are turned off.
This tests local duality only for the charged current part of Eq.~(\ref{HW}).  

The charged current diagram Fig.~\ref{fig:2}(a) factorizes in the large $N_c$
limit, so Eq.~(\ref{rate}) can be used. Combining this with
Eqs.~(\ref{womeg}), (\ref{tauspec}), and (\ref{slrate}), one finds 
\begin{eqnarray}\label{pred}
{ {\rm d} \gnlB \over {\rm d} q^2 } &=& \abs{ { \cnl \over \csl \ct}}^2
  { {\rm d} \gslB \over {\rm d}q^2 }\,
  {{\ot_T} (q^2) \over {\oslB_T} (q^2)}, \nonumber \\*
{{\ot_T} (q^2) \over {\oslB_T} (q^2)} &=& {{\rm d}\ght \over {\rm d}q^2}\,
  {96\pi^3 m_\tau^3\over G_F^2}\, {1\over(m_\tau^2-q^2)^2\,(m_\tau^2+2q^2)}\,.
\end{eqnarray}
$\Omega_L$ has been neglected in this equation because it is proportional to
the current quark masses of the $u$, $d$ and $s$ quarks. In the large $N_c$
limit, one can compute the nonleptonic $B$ decay rate in terms of the
semileptonic $B$ decay rate and the $\tau$ hadronic decay rate. This is the
main result in this paper.

Equation~(\ref{pred}) will be used to test local duality in nonleptonic $B$
decay. Assuming that the OPE computation of the total semileptonic decay rate
is valid, one knows that the integral of the experimentally measured spectrum
${\rm d} \gslB / {\rm d}q^2$ over $q^2$ is equal to the OPE value of the total
decay rate. Moments of the measured $q^2$ spectrum and of the OPE prediction
should also agree; however, the higher order $\alpha_s$ and $1/m$ corrections
get larger for higher moments. As a result, the integral of the $q^2$ spectra
from experiment and from the OPE should agree when weighted with a sufficiently
smooth and broad function, but not necessarily point by point. A similar result
holds for the $q^2$ spectrum in $\tau$ hadronic decay. In Eq.~(\ref{pred})
above, the experimentally measured $q^2$ spectrum in semileptonic $B$ decay is
multiplied by the experimentally measured $q^2$ spectrum in hadronic $\tau$
decay. The $\tau$ decay spectrum is not a smooth function (see Fig.~3), which
is what makes the test non-trivial.

\subsection{Error estimates}

The test of local duality is to compare $\gnlb$ in Eq.~(\ref{bcu}) calculated
using the OPE and perturbation theory  with $\gnlB$ in Eq.~(\ref{pred})
obtained from experimental data. Before we discuss the result, it is worth
examining the errors in the formul\ae\ we have derived. The leading corrections
to Eq.~(\ref{pred}) are factorization violating diagrams involving  two-gluon
exchange between the two quark loops in Fig.~\ref{fig:2}(a). Single gluon
exchange is forbidden by color conservation. If both gluons are hard, the
diagram can be computed in perturbation theory. It is  order $\alpha_s^2$, but
is not part of the BLM series\cite{BLM} $(\alpha_s/\pi)^{r+1}\beta_0^r$, where
$\beta_0=11-2n_f/3$  is the coefficient of the one-loop QCD $\beta$-function.
If both gluons are soft, the correction is of order $\lqcd^4/m_b^4$, and if one
gluon is hard and one is soft, the correction is of order $\alpha_s
\lqcd^2/m_b^2$. The leading power suppressed corrections to Eq.~(\ref{pred})
are of order $\alpha_s\Lambda_{\rm QCD}^2/m_b^2$ or $\Lambda_{\rm
QCD}^3/m_b^3$. The $\lqcd^2/m_b^2$ corrections cancel between $\gnlB$ and
$\gslB$. The first correction in the OPE for $\Pi_{\mu\nu}$ is order
$\Lambda_{\rm QCD}^4/q^4$. Both $\tau$ decay and $B$ decay involve summing over
$\bar d u$ and  $\bar s u$ final states with weights $\abs{V_{ud}}^2$ and
$\abs{V_{us}}^2$, respectively, so both contributions are correctly included in
Eq.~(\ref{pred}) when one uses the measured $\tau$ hadronic decay rate. In a
world with $c_2=0$, the largest correction to Eq.~(\ref{pred}) is the Pauli
interference contribution,\footnote{It is possible that the non-BLM
$\alpha_s^2$ correction, which has not yet been computed, is large.} of order
$(16 \pi^2/N_c) f_B^2/m_B^2 \sim 5\%$. Very little use has been made of the
$1/N_c$ expansion in the theoretical analysis for $c_2=0$. However, the skeptic
might argue that the error estimate is based on the OPE, which is what is being
tested in the first place. The factorization violation contributions to the
decay rate are $1/N_c^2$ corrections. This gives an  OPE-independent error
estimate of $\sim 10\%$.

\subsection{Data analysis}

The hadronic $\tau$ decay spectrum (i.e., $\ot$) is only measured in the region
$q^2<m_\tau^2$. In the region $\Delta^2 < q^2 < m_B^2$, (with $\Delta^2 <
m_\tau^2$), we use the perturbative calculation~\cite{BNP}, including
corrections to $\ot(q^2)$ up to order $\alpha_s^2$ with $\alpha_s (m_\tau ) =
0.32$. Higher order terms are negligible for our purposes. We vary $\Delta^2$
between $2-3\,{\rm GeV}^2$ and check that our conclusions are unaffected by the
precise value of $\Delta$. The region $q^2<m_\tau^2$ is where the largest
violations of local duality are expected in $\Omega$, and experimental data
suggests that local duality already holds at some level for
$q^2\gtrsim1.5\,{\rm GeV}^2$. For $W_{\mu\nu}$ the largest deviations from
local duality are expected near $q^2_{\rm max}\simeq(m_b-m_c)^2$.  

\begin{figure}
\centerline{\epsfysize=7.5truecm \epsfbox{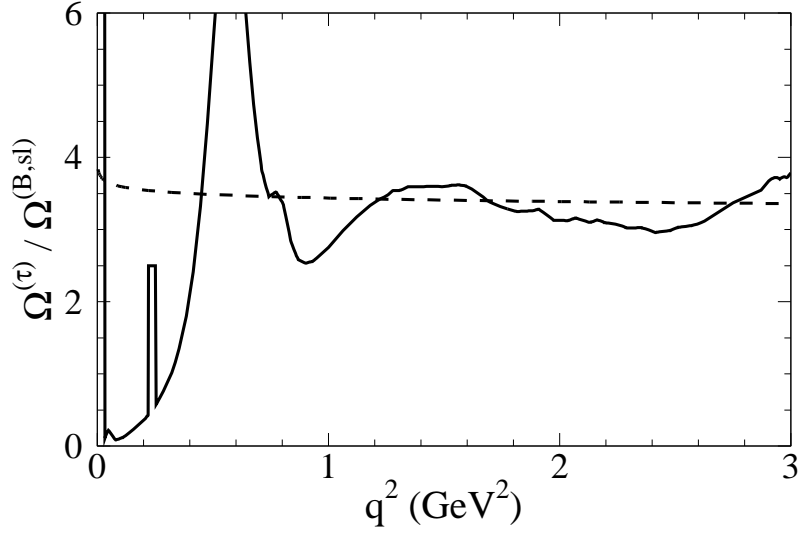} }
\caption{$\ot_T(q^2)/\oslB_T(q^2)$ from Monte Carlo fitted to $\tau$ decay data
(solid curve), and the prediction of perturbation theory  (dashed curve). The
large resonance peak is due to the $\rho$. \label{figtau} }
\end{figure}

\begin{figure} 
\centerline{\epsfysize=7.5truecm \epsffile{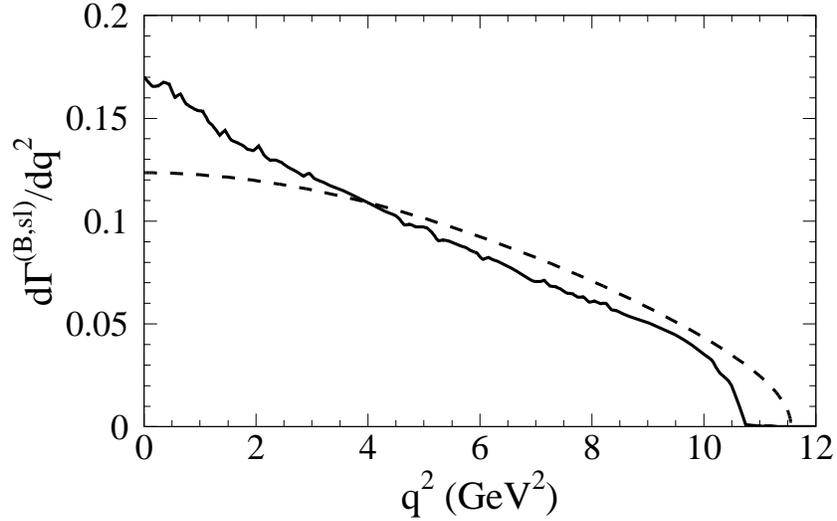} }
\caption{Monte Carlo (solid  curve) and perturbation theory (dashed curve)
predictions for $(1/\gslB)\,{\rm d}\gslB /{\rm d}q^2$. The area under each
curve has been normalized to unity. \label{figBsl} } 
\end{figure}

For the hadron invariant mass spectrum in $\tau$ decay we use a Monte Carlo
distribution that has been tuned to be consistent with CLEO data.\footnote{We
thank A.~Weinstein for the Monte Carlo spectrum.  Our results would not change
if we used the ALEPH measurement \cite{ALEPH}.}  This can be used to determine
$\ot_T(q^2)$. In Fig.~\ref{figtau} we have plotted $\ot_T(q^2)/\oslB_T(q^2)$
together with the perturbation theory prediction for this ratio (including
corrections up to order $\alpha_s^2$).  The lepton invariant mass spectrum in
semileptonic $B$ decay has not yet been measured, so we use the CLEO Monte
Carlo.\footnote{We thank E.~Potter and A.~Weinstein for the Monte Carlo
spectra.} The normalized mass spectrum is plotted in Fig.~\ref{figBsl} as a
function of $q^2$ together with the prediction of perturbation theory. The
shape of this curve is rather insensitive to the order $\alpha_s$ and
$\alpha_s^2\beta_0$ corrections\cite{LSW}.

It is simplest to normalize the nonleptonic rate to the semileptonic rate. 
This eliminates the order $\Lambda_{\rm QCD}^2/m_b^2$ corrections and reduces
the perturbative corrections as well. The result for the $B$ meson decay rate
using the factorization formula Eq.~(\ref{pred}) and the CLEO Monte Carlo
spectra is
\begin{equation}\label{finalexp}
\gnlB [b\to c \bar u(d+s)] / \gslB \bigg|_{\rm CLEO} = 3.32 \abs{c_1}^2.
\end{equation}
The result for the OPE calculation at this order is
\begin{equation}
{{\rm d}\gnlb [b \to c \bar u (d+s)]/{\rm d}q^2 \over 
  {\rm d} \gslb /{\rm d}q^2} = 3\abs{c_1\, \cnl \over \csl}^2
   \left[ 1 + {\alpha_s(q^2)\over\pi}
  + 0.173\beta_0 \left({\alpha_s(q^2)\over\pi}\right)^2
  + \ldots \right] \,.
\end{equation}

Expressing $\alpha_s(q^2)$ in terms of $\alpha_s(m_b)=0.22$ and neglecting
nonperturbative and higher order perturbative corrections, this implies  $\gnlb
[b\to c \bar u(d+s)] / \gslb = 3 \abs{c_1 \cnl/\csl}^2  \left (1 + 0.07 + 0.025
\right) = 3.21 \abs{c_1}^2$.  The difference is surprisingly small, at the 3\%
level, which is less than the 5\% theoretical uncertainty of the  calculation.
So we conclude that the uncertainties in the charged current contribution to
$b$ hadron lifetimes due to local duality violation are unlikely to exceed the
$\sim 5\%$ percent level. Of course, it is very important to repeat this
analysis using real data for the lepton mass spectrum in semileptonic $B$
decay, ${\rm d}\gslB /{\rm d}q^2$, especially since the Monte Carlo predictions
used above (solid curve in Fig.~\ref{figBsl}) differ significantly from
perturbation theory in the low $q^2$ region.

\section{Reality}

In the real world with $c_2\neq0$, there are several effects that restrict our
calculational ability. The interference term Fig.~\ref{fig:2}(c), is a $1/N_c$
correction and does not factorize.  Its contribution is order $(2/N_c)|c_2/c_1|
\sim 18\%$ of the $B \to c \bar u (d+s)$ decay rate. It has been argued that
deviations from local duality are likely to be larger for this
contribution\cite{Shifman}.

The neutral current decay Fig.~\ref{fig:2}(b) also factorizes. $W_{\mu\nu}$ for
this decay can be determined from semileptonic $B$ decay due to the
(hypothetical) neutral current $\bar d \gamma^\mu P_L b$. This can be related
to semileptonic $B$ decay due to the weak current $\bar u \gamma^\mu P_L b$ by
isospin invariance. $\Omega_{\mu\nu}$ for the neutral current decay is the
vacuum polarization contribution due to the $\bar c \gamma^\mu P_L u$ current,
for which there is no experimental information. The neutral current
contribution to the decay width is order $|c_2/c_1|^2 \sim 7 \%$.

We have neglected $b \to c \bar c (s+d)$ decays. These can also be factorized
in the large $N_c$ limit. The vacuum polarization $\Omega$  is due to $\bar s
\gamma^\mu P_L c$ and $\bar d \gamma^\mu P_L c$ currents, for which there is no
experimental information. Thus our calculational method, while valid in
principle, is not useful for $b \to c \bar c (s+d)$. Experimentally, one can
separate the $b \to c \bar u (s+d)$ and $b \to c \bar c (s+d)$ contributions to
the nonleptonic decay rate by counting the number of charm quarks in the final
state, and the results of the previous section can be used for the $b \to c
\bar u (s+d)$ part.

\section{Conclusions}

We have estimated the violation of local duality for nonleptonic $B$ decay with
$c_2=0$ to be less than the theoretical uncertainty in our calculation of $\sim
5\%$. It is important to redo the analysis using experimental data, rather than
Monte Carlo, once it is available. One can also view the results in a different
way: the factorization formula allows one to compute the nonleptonic $B$ decay
rate using the lepton mass spectrum in semileptonic $B$ decay and the hadron
mass spectrum in $\tau$ decay without using an OPE. In the world $c_2=0$, one
can predict this rate with an accuracy of 5\%. In the real world, one can
compute the $b \to c \bar u(s+d)$ part of the decay rate with approximately
20\% accuracy. The results cannot easily be extended to $D$ meson decays,
because the $\lqcd^3/m_c^3$ corrections in the charm sector are $\sim 100\%$ of
the charm quark decay result.  

These reasonably tight limits on local duality violation do not resolve the
question of the $\Lambda_b$ to $B$ lifetime ratio.  Of course, duality
violation in $b \to c \bar c (d+s)$ and in the $c_1c_2$ interference
contributions are probably larger.  While enhancing the average number of charm
in $B$ decay seems experimentally disfavored, the $c_1c_2$ interference remains
suspect.  It is also possible that spectator effects are anomalously large in
$\Lambda_b$ decay---in fact, there are six-quark operators (see Fig.~5) which
contribute roughly of order $(16\pi^2)^2(f_B/m_b)^4/N_c$ relative to the $b$
quark decay.  These have not been analyzed, and may significantly affect the
spectator contribution to $\Lambda_b$ decay.  They do not occur for $B$ meson
decay, and so contribute to the $\Lambda_b-B$ lifetime difference.

In semileptonic $B$ decay, there is no $1/m_b$ correction~\cite{CGG}. This is
also true in nonleptonic $B$ decays, if one uses the OPE~\cite{VoSi}.  In other
processes which do not have an OPE, it is known that one can get arbitrary
power suppressed corrections~\cite{Wise}. It has been suggested that there
might be $1/m_b$ corrections in inclusive nonleptonic decay rates~\cite{Lebed}.
In the large $N_c$ limit with $c_2=0$, Eq.~(\ref{pred}) shows that the $1/m_b$
corrections to the nonleptonic decay rate can be computed in terms of the
corrections to the lepton mass spectrum in semileptonic $B$ decay and the
hadron mass spectrum in $\tau$ decay. At least one of these would have to have
a $1/m_b$ correction if there is one in the nonleptonic decay rate.

\acknowledgements  
We thank E.~Potter and A.~Weinstein for providing us with
Monte Carlo simulations, M.E.~Luke for some numerical results, and M.~Gremm,
V.~Sharma, I.~Stewart, and M.B.~Wise for discussions.  This work was supported
in part by the Department of Energy under grant DOE-FG03-97ER40546, and by the
National Science Foundation under grant PHY-9457911.

\newpage

\begin{figure}
\centerline{\epsfysize=3cm\epsffile{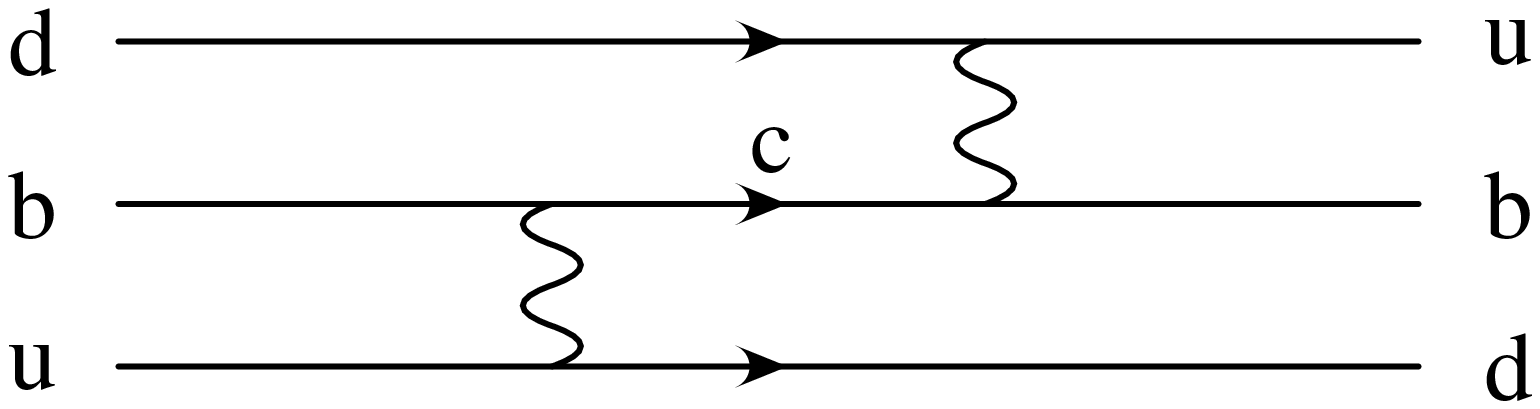}}
\caption{Spectator diagram which contributes to the $\Lambda_b$ lifetime, 
but has no analog for $B$ meson decay. 
\label{fig:5}}
\end{figure}


\begin{references}

\bibitem{PQW}
E.C. Poggio, H.R. Quinn, and S. Weinberg, Phys. Rev. D13 (1976) 1958.

\bibitem{boyd} 
C.G.~Boyd, B.~Grinstein, and A.V.~Manohar, Phys.\ Rev.\ D54 (1996) 2081.

\bibitem{CGG}
J. Chay, H. Georgi, and B. Grinstein, Phys. Lett. B247 (1990) 399.

\bibitem{VoSi}
M. Voloshin and M. Shifman, Sov. J. Nucl. Phys. 41 (1985) 120.

\bibitem{BSUV}
I.I. Bigi, M. Shifman, N.G. Uraltsev, and A. Vainshtein,
Phys. Rev. Lett. 71 (1993) 496.

\bibitem{MaWi}
A.V. Manohar and M.B. Wise, Phys. Rev. D49 (1994) 1310.

\bibitem{Blok}
B. Blok, L. Koyrakh, M. Shifman, and A.I. Vainshtein, 
Phys. Rev. D49 (1994) 3356.

\bibitem{Mannel}
T. Mannel, Nucl. Phys. B413 (1994) 396.

\bibitem{inclnl}
I.I. Bigi, N.G. Uraltsev, and A.I. Vainshtein, Phys. Lett. B293 (1992) 430
[(E) Phys. Lett. B297 (1993) 477]; 
B. Blok and M. Shifman, Nucl. Phys. B399 (1993) 441; 459.

\bibitem{tHooft}
G. 't Hooft, Nucl. Phys. B72 (1974) 461; Nucl. Phys. B75 (1974) 461.

\bibitem{BuBu}
G. Buchalla and A.J. Buras, Nucl. Phys. B400 (1993) 225.

\bibitem{GNPR}
B. Guberina, S. Nussinov, R.D. Peccei, and R. Ruckl, 
Phys. Lett. B89 (1979) 111.

\bibitem{BGR}
A.J. Buras, J.-M. Gerard, and R. Ruckl, Nucl. Phys. B268 (1986) 16.

\bibitem{BNP}
E. Braaten, S. Narison, and A. Pich, Nucl. Phys. B373 (1992) 581
and references therein.

\bibitem{FLNN}
A.F. Falk, Z. Ligeti, M. Neubert, and Y. Nir, Phys. Lett. B326 (1994) 145; 
L. Koyrakh, Phys. Rev. D49 (1994) 3379; S. Balk, J.G. Korner, D. Pirjol, and
K. Schilcher, Z.~Phys. C64 (1994) 37.

\bibitem{BLM} 
S.J.~Brodsky, G.P.~Lepage, and P.B.~Mackenzie, Phys.\ Rev.\ D28 (1983) 228.

\bibitem{ALEPH}
R. Barate {\it et al.}, ALEPH Collaboration, CERN PPE/98-012.

\bibitem{LSW} 
M.E. Luke, M.J. Savage, and M.B. Wise,  Phys. Lett. B343 (1995) 329. 

\bibitem{Shifman} 
M.A. Shifman, Nucl. Phys. B388 (1992) 346.

\bibitem{Wise} 
A.V.~Manohar and M.B.~Wise, Phys.\ Lett.\ B344 (1995) 407.

\bibitem{Lebed} 
B.~Grinstein and R.F.~Lebed, Phys.\ Rev.\ D57 (1998) 1366.

\end{references}
\end{document}